\documentclass[12pt]{article}
\author{N.V.\,Ustinov \\
\it Department of Theoretical Physics, \\
\it Kaliningrad State University, \\
\it Al.\,Nevsky street 14, Kaliningrad, 236041, Russia }
\title{ Darboux Transformations, Infinitesimal Symmetries and Conservation
Laws for Nonlocal Two--Dimensional Toda Lattice}
\date{ }
\begin{document}
\maketitle

{\bf Short title:} Nonlocal Two--Dimensional Toda Lattice

{\bf Classification numbers:}  PACS 05.45.Yv, 02.30.Jr

\begin{abstract}
The technique of Darboux transformation is applied to nonlocal partner of
two--dimensional periodic $A_{n-1}$ Toda lattice.
This system is shown to admit a representation as the compatibility conditions
of direct and dual overdetermined linear systems with quantized spectral
parameter.
The generalization of the Darboux transformation technique on linear equations
of such a kind is given.
The connections between the solutions of overdetermined linear systems and
their expansions in series at singular points neighborhood are presented.
The solutions of the nonlocal Toda lattice and infinite hierarchies of the
infinitesimal symmetries and conservation laws are obtained.
\end{abstract}

\pagebreak
\section{Introduction}
It is well known that some of nonlocal differential equations of great
physical significance (e.g., the Benjamin--Ono (BO) equation
\cite{1}--\cite{3}, the intermediate long wave (ILW) equation \cite{4,5})
possess the same mathematical properties as local nonlinear equations
integrable in the frameworks of the inverse scattering transformation (IST)
method \cite{6}.
Such the nonlocal equations were found to have the multi--soliton solutions
\cite{4},\cite{7}--\cite{11},
infinitely many conservation laws \cite{12},\cite{15}--\cite{18},
the B\"acklund transformation \cite{15}--\cite{18},
to pass the Painlev$\acute{\mbox{\rm e}}$ test \cite{18},
to be represented in the bilinear form \cite{8}--\cite{11},%
\cite{15}--\cite{18} and as the compatibility condition of the overdetermined
linear system (Lax pair) \cite{13}--\cite{18},
to possess the algebro--geometric solutions \cite{19}.
An effective technique of Darboux transformation (DT) \cite{20} was also
applied to nonlocal equations, such as the ILW equation, nonlocal analogue of
the Kadomtsev--Petviashvili equation \cite{14,19} and nonlocal Toda equation
\cite{21}.
This technique allows one to obtain the infinite hierarchies of solutions of
Lax pairs and associated nonlinear equations to arise from the compatibility
condition.
Carrying out the proper sequence of the Darboux transformations adds the 
one--soliton component of solution to initial solution of nonlinear integrable 
equations.

The noncommutative ("quantum") generalization of the spectral parameter of
usual Lax pair was suggested in \cite{22} to construct the hierarchies of 
nonlocal counterparts of the nonlinear equations admitting the compatibility 
condition representation.
This way the hierarchies of the ILW$_n$ equations, modified ILW$_n$ equations
and nonlocal Toda lattice were obtained \cite{22,23}.
Nonlocal bilinear equations for the last system, which were proposed in work
\cite{24}, should be modified for the multi--soliton solutions to be derived 
by applying the ordinary procedure of the perturbation theory to a constant
solution \cite{25}.
It is also inconvenient in the frameworks of the bilinear approach to
determine a set of exponents, whose mutual products appear no in the expansion
for the multi--soliton solutions, and, consequently, to describe the soliton 
solutions.

In the present paper we extend the DT technique to nonlocal partners of Lax
pairs and associated nonlocal nonlinear equations.
The dual Lax pair with quantized spectral parameter is introduced in Sec.II.
The connections between the spaces of solutions of direct and dual Lax pairs
are found.
The reduction constraint on the Lax pairs coefficients that leads to nonlocal
$A_{n-1}$ Toda lattice $(n\in\bf N)$ is discussed in Sec.III.
This nonlinear system is written here in the bilinear form, which is suitable
for exploiting the usual perturbation theory and whose one--soliton expansion,
as it is shown in the next section, has infinitely many exponential terms.
In Sec.IV the theorem establishing the covariance of the Lax pairs with
quantized spectral parameter with respect to the DT of direct pair is
presented.
For the case of the Toda lattice, we formulate sufficient conditions to keep
the reduction constraint on the Lax pair coefficients under performing this 
transformation.
Similarly we construct the DT of dual pair preserving the reduction, whose
product with the DT of direct pair yields the formulas of binary DT.
Iterations of these transformations on zero background give multi--soliton
solutions that depend on infinitely many arbitrary parameters.
The formulas of infinitesimal Darboux transformation and the expansions in
series of the Lax pair solutions at the singular points neighborhood are 
presented in Sec.V.
A simple way to produce infinite hierarchies of the infinitesimal symmetries
and conservation laws of nonlocal $A_{n-1}$ Toda lattice is given there.

\section{Lax Pairs with Quantized Spectral Pa\-ra\-me\-ter}
Let us consider the overdetermined linear system
\begin{equation}
\left\{
\begin{array}{l}
\Psi_x = - J \, T \Psi \Lambda + U\Psi \\
\Psi_t = A \, T^{-1} \Psi \, \Lambda^{-1}
\end{array}
\right.
\label{1}
\end{equation}
for matrix $n * n$ function \mbox{$\Psi \equiv \Psi (x, t, \Lambda)$}.
Here $\Lambda = \mbox{diag} (\lambda_j)$ is the constant matrix,
$T$ is the translation operator:
$$
T = \exp ( h \partial_x )
$$
($h$ is a constant), matrices $J$, $U$ and $A$ are independent of $\Lambda$.
This system coincides with usual Lax pair studied in the frameworks of the IST
method \cite{6} if $h=0$.
Matrix $\Lambda$ in this case is called the matrix spectral parameter.
Joint action of operator $T$ and matrix $\Lambda$ on solution $\Psi$ can be
regarded as a notion of the quantized spectral parameter \cite{22}.
We will quote Eqs.(\ref{1}) as direct Lax pair in that follows.
Compatibility condition $\Psi_{x\,t}=\Psi_{t\,x}$ of direct Lax pair gives
system of nonlocal matrix equations
\begin{equation}
\left\{
\begin{array}{l}
J_t = 0 \\
U_t = J \, T A - A \, T^{-1} J \\
A_x = UA - A\,T^{-1}U \\
\end{array}
\right. .
\label{2}
\end{equation}

It is remarkable that Eqs.(\ref{2}) are also derived from the compatibility
condition of the overdetermined linear system that is "dual" to Lax pair
(\ref{1}).
This system (dual Lax pair) is written in following manner
\begin{equation}
\left\{
\begin{array}{l}
\Xi_x = K \left( T^{-1} \Xi J \right) - \Xi U \\
\Xi_t = - K^{-1} \left( T \Xi A \right)
\end{array}
\right.\!\!
\label{3}
\end{equation}
with $\Xi \equiv \Xi (x, t, K)$ and $K = \mbox{diag} (\symbol{26}_j)$ being
respectively matrix solution and matrix spectral parameter of dual system.
The spaces of solutions of direct and dual Lax pairs are connected in the
local case.
Such the connection turns out to exist for the Lax pairs considered here.
Namely, matrix function
$$
R (\Xi,\Psi) = K \int_x^{x+h} \left( T^{-1} \Xi J \right) \Psi dx + \Xi\Psi
$$
is independent on variables $x$ and $t$ if $K = \Lambda = \lambda E $
($\lambda$ is scalar spectral parameter).
Another relation between the spaces of solutions is provided by the closure of
differential 1--form
\begin{equation}
d \omega (\Xi,\Psi) = \Xi J T \Psi dx + K^{-1} \left( T \Xi A \right) \Psi
\Lambda^{-1} dt .
\label{dw}
\end{equation}

\section{Nonlocal Two--Dimensional Toda Lattice}
Let matrix $J$ be defined as given
\begin{equation}
J = \{\delta_{j,k-1}\}
\label{J}
\end{equation}
(the indices are supposed hereafter to be equal on modulo $n$).
System (\ref{2}) is valid if we put
\begin{equation}
U = \sigma_x , \,\,\,
A = \exp (\sigma) J^{-1} \exp(-T^{-1} \sigma) ,
\label{4}
\end{equation}
where
\begin{equation}
\sigma = \mbox{\rm diag} (\sigma_j)
\label{sigma}
\end{equation}
and functions $\sigma_j \equiv \sigma_j (x,t)$ ($j=1,...,n$) are the solutions
of nonlocal generalization of the two--dimensional periodic $A_{n-1}$ Toda
lattice \cite{26}
\begin{equation}
\sigma_{j,xt} = \exp \left( \vphantom{T^{-1}} T \sigma_{j+1} - \sigma_j \right)
- \exp \left( \sigma_j - T^{-1} \sigma_{j-1} \right) .
\label{5}
\end{equation}
Changing variables
$$
\sigma_j = \log \frac{T \tau_{j+1}}{\tau_j}
$$
results Eqs.(\ref{5}) in bilinear form:
\begin{equation}
D_x D_t \tau_j \cdot \tau_j + 2 \tau_j^2 =
2 \left( T \tau_{j+1} \right) T^{-1} \tau_{j-1}
\label{6}
\end{equation}
($D_x$ and $D_t$ denote the Hirota derivatives \cite{25}).
In the local case these equations give the bilinear representation of the
two--dimensional Toda lattice (compare with Eqs.(4.3) in \cite{24}).

The important feature of nonlocal direct and dual Lax pairs is that matrix
$U$ of general form cannot be led to the algebra $SL(n)$ case by means of the
gauge transformation \cite{6}.
However, the coefficients of linear systems (\ref{1},\ref{3}) admit new reduction constraint that is in keeping with system (\ref{5}):
\begin{equation}
\sigma = \rho E ,
\label{7}
\end{equation}
where single dependent variable $\rho$ solves equation
($A_0$ Toda lattice)
\begin{equation}
\rho_{xt} = \left( T - 1 \right) \exp \left( \rho - T^{-1} \rho \right) .
\label{8}
\end{equation}
It should be stressed that this equation, whose local analogue is trivial,
appears as the compatibility condition of the overdetermined linear systems of
arbitrary matrix dimension.
Making the dependent variable transformation in Eq.(\ref{8})
$$
\rho = (T-1) \log \tau ,
$$
one obtains bilinear equation for $\tau$
\begin{equation}
D_x D_t \tau \cdot \tau + 2 \tau^2 = 2 \left( T \tau \right) T^{-1} \tau.
\label{9}
\end{equation}
Eq.(\ref{8}) was derived in \cite{B1} as continious limit of the lattice 
equations describing the transfer of energy of plasma oscillations.
The nonlocal equation that can be written in the form of Eq.(\ref{8}) with
operator $T$ defined in different manner was considered in work \cite{21}.

The soliton solutions of Eqs.(\ref{6}) and Eq.(\ref{8}) can be constructed 
applying the usual procedure of the Hirota method \cite{25}.
However, the explicit form of the one--soliton solution is not obvious in this
approach.
To generate the hierarchy of solutions of the nonlocal Toda lattice we develop
below another method.

\section{Darboux Transformation Technique and Solitons}
The underlying property of the DT technique is an existence of the kernel of
transformations of the Lax pairs solutions for some value of the spectral
parameter.
This property is exploited in the present paper to generalize the technique
considered for the nonlocal equations.
For the sake of convenience we introduce notation
$$
\Omega(\Xi,\Psi) = \int_{(x_0,t_0)}^{(x,t)} d \omega(\Xi,\Psi) + C(\Xi,\Psi) ,
$$
where it is supposed that constant matrix $C(\Xi,\Psi)$ can be determined from
equation
$$
K C(\Xi,\Psi) - C(\Xi,\Psi)\Lambda = \left.
R ( \Xi , \Psi ) \right|_{(x_0,t_0)} .
$$

{\sc THEOREM} {\it Let $\Phi$ be a solution of direct Lax pair (1) with matrix
spectral parameter $M$. Then matrix
\begin{equation}
\tilde \Psi = \Psi_x - \Phi_x \Phi^{-1} \Psi
\label{10}
\end{equation}
and, if there exists appropriate matrix $C(\Xi,\Phi)$, matrix
\begin{equation}
\tilde \Xi = \Omega(\Xi,\Phi) \left( T\Phi^{-1} \right) J^{-1}
\label{11}
\end{equation}
are solutions respectively of direct and dual Lax pairs
$$
\left\{
\begin{array}{l}
\tilde \Psi_x = - J \, T \tilde \Psi \Lambda + \tilde U\tilde \Psi \\
\tilde \Psi_t = \tilde A \, T^{-1} \tilde \Psi \, \Lambda^{-1}
\end{array}
\right. , \,\,\,\,\,\,\,
\left\{
\begin{array}{l}
\tilde \Xi_x = K \, \left( T^{-1} \tilde \Xi J \right) - \tilde \Xi \tilde U \\
\tilde \Xi_t = - K^{-1}\, \left( T \tilde \Xi \tilde A \right)
\end{array}
\right. ,
$$
whose coefficients are
\begin{equation}
\tilde U = U + J_x J^{-1} + J \left( T \Phi_x\Phi^{-1} \right) J^{-1} - \Phi_x\Phi^{-1} ,
\label{12}
\end{equation}
\begin{equation}
\tilde A = J  (T\Phi) M \Phi^{-1} A (T^{-1}\Phi) M^{-1} \Phi^{-1} (T^{-1}J^{-1}) .
\label{13}
\end{equation}
}

The direct Lax pair is covariant with respect to transformation
$\{ \Psi, U , A \} \to \{ \tilde \Psi, \tilde U, \tilde A \}$ owing to the
existence of the kernel: $\tilde \Psi \equiv 0$ if $\Psi = \Phi$.
The covariance of the dual Lax pair is proven by applying identity
$$
K \left( T^{-1} \Omega (\Xi,\Phi) \right) = \Omega (\Xi,\Phi) M + \Xi \Phi .
$$
It follows from the compatibility conditions of transformed direct or dual
Lax pairs that their coefficients $J$, $\tilde U$ and $\tilde A$ are new
solutions of system (\ref{2}).
Transformation (\ref{10}--\ref{13}) carried out with solution $\Phi$ of
Eq.(\ref{1}) is called the DT of direct pair.
Reduction constraint (\ref{4}) on the Lax pairs coefficients is kept under
performing this DT by imposing the following conditions on solution
$\Phi$ and its spectral parameter $M$:
\begin{equation}
\Phi = \mbox{\rm diag} (\varphi_j) B ,
\label{14}
\end{equation}
$$
B_{jk}= \exp \,(2\pi i (j-1)(k-1)/n)
$$
$$
(j,k=1,...,n) ,
$$
\begin{equation}
M = \mu \, \mbox{\rm diag} ( \exp \, (2\pi i (j-1)/n) ) .
\label{15}
\end{equation}
Here $\varphi \equiv \varphi (x,t,\mu) = (\varphi_1,...,\varphi_n)^T$ is
a vector solution of Eq.(\ref{1}) with scalar spectral parameter $\mu$.
If $\psi = (\psi_1,...,\psi_n)^T$ and $\xi = (\xi_1,...,\xi_n)$ are the vector
solutions of direct and dual Lax pairs with scalar spectral parameters
$\lambda$ and $\symbol{26}$ respectively, then Eqs.(\ref{10},\ref{11}) under
conditions (\ref{14},\ref{15}) give the next expressions for the components
of transformed vector solutions
\begin{equation}
\displaystyle \tilde \psi_j = \psi_{j,x} -
\frac{\varphi_{j,x}}{\varphi_j} \psi_j ,
\label{16}
\end{equation}
\begin{equation}
\displaystyle \tilde \xi_j = \frac{\Omega_{j+1}(\xi,\varphi)}{T\varphi_{j+1}}
\label{17}
\end{equation}
$(j=1,...,n)$ with
$$
\displaystyle \Omega_j(\xi,\varphi) =
\int_{(x_0,t_0)}^{(x,t)} \xi_{j-1} T \varphi_j \, dx +
\exp \left( T \sigma_j - \sigma_{j-1} \right)
\frac{\left( T \xi_j \right) \varphi_{j-1}}{\symbol{26}\mu} dt 
+\left.\frac{\rho_j(\xi,\varphi)}{\symbol{26}^n-\mu^n} \right|_{(x_0,t_0)} ,
$$
$$
\rho_j(\xi,\varphi) = \sum_{k=1}^n \symbol{26}^{k-1} \mu^{n-k}
\left( \symbol{26} \int_x^{x+h} \left( T^{-1}\xi_{j-k-1} \right)
\varphi_{j-k} dx + \xi_{j-k} \, \varphi_{j-k} \right) .
$$
Transformation of solutions of nonlocal Toda lattice (\ref{5}), which
corresponds to transformations (\ref{16},\ref{17}) of solutions of associated
Lax pairs, has the form
\begin{equation}
\displaystyle \tilde \sigma_j = \sigma_j +
\log \frac{T\varphi_{j+1}}{\varphi_j} \,\, (j=1,...,n) .
\label{18}
\end{equation}

Similarly one can construct the formulas of DT of dual pair, using the matrix
solution of Eq.(\ref{3}).
It can be shown that the transformations of direct and dual pairs commute.
The conditions analogous to Eqs.(\ref{14},\ref{15}) are imposed on the matrix
solution and its spectral parameter for reduction
constraint (\ref{4}) to be inherited under carrying out DT of dual pair.
Corresponding formulas for the transformations of the Lax pairs solutions and 
ones of nonlocal Toda lattice are
\begin{equation}
\displaystyle \tilde \psi_j = T^{-1}
\frac{\Omega_{j}(\chi,\psi)}{\chi_{j-1}} ,
\label{19}
\end{equation}
\begin{equation}
\displaystyle \tilde \xi_j = \xi_{j,x} -
\frac{\chi_{j,x}}{\chi_j} \xi_j ,
\label{20}
\end{equation}
\begin{equation}
\displaystyle \tilde \sigma_j = \sigma_j + \log \frac{\chi_j}{T^{-1}\chi_{j-1}}
\label{21}
\end{equation}
$(j=1,...,n)$,
where $\chi=(\chi_1,...,\chi_n)$ is the vector solution of dual Lax pair with
scalar spectral parameter $\nu$.
The product of transformations (\ref{16}--\ref{18}) and (\ref{19}--\ref{21})
yields the formulas of so--called binary DT
\begin{equation}
\displaystyle \tilde \psi_j = \psi_j -
\frac{T^{-1}\Omega_j(\chi,\psi)}{T^{-1}\Omega_j(\chi,\varphi)}\varphi_j ,
\label{22}
\end{equation}
\begin{equation}
\displaystyle \tilde \xi_j = \xi_j -
\frac{\Omega_{j+1}(\xi,\varphi)}{\Omega_{j+1}(\chi,\varphi)}
\chi_j ,
\label{23}
\end{equation}
\begin{equation}
\displaystyle \tilde \sigma_j = \sigma_j +
\log \frac{\Omega_{j+1}(\chi,\varphi)}{T^{-1}\Omega_j(\chi,\varphi)}
\label{24}
\end{equation}
$(j=1,...,n)$.

To keep reduction constraint (\ref{7}) we have to impose additional conditions
on the solutions of Lax pairs, which are used in performing DT's.
For example, if we put $\varphi_j = \alpha^j \vartheta$ in transformation
(\ref{16}--\ref{18}), where $\alpha = \mbox{exp} (2\pi i k /n)$ $(k=0,...,n-1)$
and function $\vartheta$ solves system
$$
\left\{
\begin{array}{l}
\vartheta_x = - \alpha \mu T \vartheta + \rho_x \vartheta \\
\displaystyle \vartheta_t =
\frac{\mbox{exp}(\rho - T^{-1}\rho)}{\alpha \mu} T^{-1} \vartheta
\end{array}
\right. ,
$$
then transformed solution of Eq.(\ref{8}) is
$$
\displaystyle \tilde \rho = \rho + (T-1) \log \vartheta .
$$

Let us consider the zero background $(\sigma=0)$.
Vectors $\varphi$ and $\chi$, which were exploited in constructing the DT's,
satisfy nonlocal linear systems
\begin{equation}
\left\{
\begin{array}{l}
\varphi_x = - \mu J \, T \varphi \\
\varphi_t = \mu^{-1} J^{-1} \, T^{-1} \varphi
\end{array}
\right.,
\label{25}
\end{equation}
\begin{equation}
\left\{
\begin{array}{l}
\chi_x = \nu \,T^{-1} \chi J \\
\chi_t = - \nu^{-1}\, T \chi J^{-1}
\end{array}
\right..
\label{26}
\end{equation}
These systems have solutions depending on infinitely many parameters of the
next form
\begin{equation}
\displaystyle\varphi_j=\sum_{k=0}^{n-1}\sum_{m_k}c_k^{(m_k)}
\exp\left(2\pi i(j-1)k/n\right)
\exp\left(p_k^{(m_k)}x-t/p_k^{(m_k)}\right)\,,
\end{equation}
\begin{equation}
\displaystyle\chi_j=\sum_{k=0}^{n-1}\sum_{n_k}d_k^{(n_k)}
\exp\left(-2\pi i(j-1)k/n\right)
\exp\left(q_k^{(n_k)}x-t/q_k^{(n_k)}\right)\,,
\end{equation}
$(j=1,...,n)$.
Here $p_k^{(m_k)}$ and $q_k^{(n_k)}$ satisfy equations
$$
p_k^{(m_k)}+\mu\exp(2\pi ik/n)\exp(p_k^{(m_k)}h)=0\,,
$$
$$
q_k^{(n_k)}-\nu\exp(2\pi ik/n)\exp(-q_k^{(n_k)}h)=0
$$
$(k=0,...,n-1)$, $c_k^{(m_k)}$ and $d_k^{(n_k)}$ are arbitrary constants.
Carrying out of DT accordingly to Eqs.(\ref{16}--\ref{18}) (or
Eqs.(\ref{19}--\ref{21})) gives the one--soliton solutions, which depend on
infinitely many free parameters.
It means that in general the set $\tau_j$ $(j=1,...,n)$ of the one--soliton
$\tau$--functions of Eqs.(\ref{6}) and the one--soliton $\tau$--functions of
Eq.(\ref{9}) contain infinitely many exponential terms.
The products of these exponents appear no in the expression of the 
multi--soliton solutions of the bilinear equations.
One--soliton solutions in nonlocal case form two families to be produced by
DT's of direct and dual pairs respectively, while in local case transformation
(\ref{19}--\ref{21}) can be obtained iterating transformation
(\ref{16}--\ref{18}).

In the case $n=2$, the components of solution of system (\ref{25}) are
represented in the following manner
$$
\displaystyle \varphi_j = \sum_{p_+} c_{p_+} \exp \left( p_+ x - t/p_+ \right)
+ (-1)^{j-1} \sum_{p_-} c_{p_-} \exp \left( p_- x - t/p_- \right)
$$
$(j=1,2)$, where $p_\pm$ are the roots of equations
$$
p \pm \mu \exp (p h) = 0 ,
$$
and the summations are supposed over all solutions of the last equations.
Substitution of vector solution $\varphi$ such, that
\mbox{$\varphi_1=\pm\varphi_2$}, into Eq.(\ref{18}) leads to the solutions of
$A_0$ Toda lattice (\ref{8}).
For real $h$, simplest real nonsingular solution is
$$
\displaystyle \rho = (T-1) \log \vartheta ,
$$
where
$$
\vartheta = c_1 \exp \left( p_1 x - t/p_1 \right)
+ c_2 \exp \left( p_2 x - t/p_2 \right) ,
$$
$c_1$, $c_2$ and $\mu$ are real constants, $c_1c_2>0$, $p_1$ and $p_2$ are
real solutions of equation
$$
p = \mu \exp (p h) .
$$

Iterations of transformations (\ref{16}--\ref{24}) allow one to construct the
infinite hierarchies of solutions of Eqs.(\ref{5}) and corresponding solutions
of Lax pairs (\ref{1},\ref{3}).
Final expressions for the transformed quantities are brought to the
determinant form.
Taking into account
$$
\lambda \rho_{j+1} (\xi,\psi) = \symbol{26} \rho_j (\xi,\psi) +
(\lambda^n - \symbol{26}^n)
\left( \symbol{26}\int_x^{x+h} \left( T^{-1}\xi_{j-1} \right) \psi_j dx +
\xi_j \, \psi_j \right) ,
$$
$$
\symbol{26} T^{-1} \Omega_j (\xi,\psi) = \lambda \Omega_{j+1} (\xi,\psi) +
\xi_j \psi_j ,
$$
one obtains
$$
\displaystyle \rho_j (\tilde \xi,\tilde \psi) = (\symbol{26}^n - \lambda^n)
\frac{T\psi_j}{T\varphi_j} \Omega_j (\xi,\varphi) - \rho_j (\xi,\psi) ,
$$
$$
\displaystyle \Omega_j (\tilde \xi,\tilde \psi) =
\frac{T\psi_j}{T\varphi_j} \Omega_j (\xi,\varphi) - \Omega_j (\xi,\psi) ,
$$
where $\tilde \psi$ and $\tilde \xi$ are defined accordingly to
Eqs.(\ref{16},\ref{17}).
The formulas of $N$-th iteration of the DT of direct pair are written by means
of these equations as given
$$
\displaystyle \tilde \psi_j = \frac{\Delta^{(j)}[N+1]}{\Delta^{(j)}[N]},
$$
$$
\tilde \xi_j =
\left|
\begin{array}{ccccc}
T \varphi^{(1)}_{j+1} & T \varphi^{(1)}_{j+1,x} & \ldots&
T \varphi^{(1)}_{j+1,(N-2)x} & \Omega_{j+1}(\xi,\varphi^{(1)}) \\
T \varphi^{(2)}_{j+1} & T \varphi^{(2)}_{j+1,x} & \ldots &
T \varphi^{(2)}_{j+1,(N-2)x} & \Omega_{j+1}(\xi,\varphi^{(2)}) \\
\vdots & \vdots & \ddots &\vdots & \vdots \\
T \varphi^{(N)}_{j+1} & T \varphi^{(N)}_{j+1,x} & \ldots &
T \varphi^{(N)}_{j+1,(N-2)x} & \Omega_{j+1}(\xi,\varphi^{(N)}) \\
\end{array}
\right|/T\Delta^{(j+1)}[N],
$$
$$
\displaystyle \tilde \sigma_j = \sigma_j +
\log \frac{T\Delta^{(j+1)}[N]}{\Delta^{(j)}[N]} ,
$$
where
$$
\Delta^{(j)}[N] = \det [ \varphi^{(k)}_{j,(m-1)x} ],
$$
$\varphi_j^{(k)}$ are the components of the vector solutions of Eq.(\ref{1})
with scalar spectral parameter $\mu^{(k)}$, $\varphi^{(N+1)}=\psi$
$(j=1,...n ; k,m =1,...,N)$.
The reduction constraint (\ref{7}) is inherited under the iterations by
supposing $N=l\,n$ $(l=1,2,...)$,
$\varphi^{(m\,l+k)} = (\varphi_{m+1}^{(k)}...,\varphi_{m+n}^{(k)})^T$,
$\mu^{(m\,l+k)} = \mu^{(k)}$ $(m=0,...,n-1;\, k=1,...,l)$.

The iterations of the transformations presented in this section on zero
background give the multi--soliton solutions of nonlocal $A_{n-1}$ Toda
lattice that depend on infinitely many arbitrary parameters.
The interaction of the one--soliton components in the multi--soliton solution
causes the shifts of the parameters of the one--soliton solutions.

\section{ Infinitesimal Symmetries and Conservation Laws }
Taking limit $\nu\to\mu$ in Eqs.(\ref{22}--\ref{24}) one finds solutions of
the linearizations of direct and dual Lax pairs and nonlocal Toda lattice
\begin{equation}
\delta\psi_j=\mu\left(T^{-1}\Omega_j(\chi,\psi)\right)\varphi_j,
\label{iDT1}
\end{equation}
\begin{equation}
\delta\xi_j=\mu\Omega_{j+1}(\xi,\varphi)\chi_j,
\label{iDT2}
\end{equation}
\begin{equation}
\delta\sigma_j=\chi_j\varphi_j
\label{iDT3}
\end{equation}
($j=1,...,n$).
These formulas establish infinitesimal DT $\psi\to\psi+\varepsilon\delta\psi$,
$\xi\to\xi+\varepsilon\delta\xi$, $\sigma\to\sigma+\varepsilon\delta\sigma$
($\varepsilon=\nu-\mu$).
The closure of differential 1--form (\ref{dw}) immediately yields
$$
(\Xi JT\Psi)_t=( K^{-1}\left(T\Xi A\right)\Psi\Lambda^{-1})_x.
$$
In terms of vector solutions $\varphi$ and $\chi$ of Lax pairs with scalar
spectral parameter $\mu$ this identity reads as
$$
\Bigl(\sum\limits_{j=1}^n\chi_{j-1}T\varphi_j\Bigr)_t=
\mu^{-2}\Bigl(\sum\limits_{j=1}^n\left(T\chi_{j+1}\right)\varphi_j\,
\mbox{e}^{T\sigma_j-\sigma_{j-1}}\Bigr)_x
$$
or, equivalently,
\begin{equation}
T_t+X_x=0,
\label{cl}
\end{equation}
where we use notations
\begin{equation}
T=\sum\limits_{j=1}^n\chi_{j-1}T\varphi_j,
\label{cd}
\end{equation}
\begin{equation}
X=-\mu^{-2}\sum\limits_{j=1}^n\left(T\chi_{j+1}\right)\varphi_j\,
\mbox{e}^{T\sigma_j-\sigma_{j-1}}.
\label{cf}
\end{equation}

The hierarchies of infinitesimal symmetries and conservation laws of nonlocal
generalization of two--dimensional periodic $A_{n-1}$ Toda lattice are
obtained substituting in Eqs.(\ref{iDT3}--\ref{cf}) the expansions of the Lax 
pairs solutions at the neighborhood of singular points on the spectral 
parameter plane.
The components of vector solutions $\varphi$ and $\chi$ of direct and dual Lax
pairs (\ref{1},\ref{3}), whose coefficients are defined by
Eqs.(\ref{J}--\ref{sigma}), are represented in the neighborhood of point
$\mu=\infty$ in the following manner
\begin{equation}
\varphi_j=\left(1+\sum\limits_{k=1}^{\infty}\frac{A_j^{(k)}}{\Omega^k}
\right)\left(\frac{\Omega}{\mu}\right)^{j-1}\,
\mbox{e}^{\displaystyle\Omega(j-1)h-\Omega x},
\label{phi_inf}
\end{equation}
\begin{equation}
\chi_j=\left(1+\sum\limits_{k=1}^{\infty}\frac{B_j^{(k)}}{\Omega^k}
\right)\left(\frac{\mu}{\Omega}\right)^{j-1}\,
\mbox{e}^{\displaystyle-\Omega(j-1)h+\Omega x}
\label{chi_inf}
\end{equation}
($j=1,...,n$), where $\Omega$ solves equation
$$
\Omega^n\mbox{exp}(n\Omega h)=\mu^n
$$
and coefficients $A_j^{(k)}$ and $B_j^{(k)}$ ($k\in\bf N$) satisfy equations
$$
\left\{
\begin{array}{l}
-A_j^{(k)}+A_{j,x_{\mathstrut}}^{(k-1)}=-TA_{j+1}^{(k)}+\sigma_{j,x}A_j^{(k-1)}\\
A_{j,t}^{(k)}=\mbox{e}^{\sigma_j-T^{-1}\sigma_{j-1}}T^{-1}A_{j-1}^{(k-1)}
\end{array}
\right.,
$$
$$
\left\{
\begin{array}{l}
B_j^{(k)}+B_{j,x_{\mathstrut}}^{(k-1)}=T^{-1}B_{j-1}^{(k)}-\sigma_{j,x}B_j^{(k-1)}\\
B_{j,t}^{(k)}=-\mbox{e}^{T\sigma_{j+1}-\sigma_j}TB_{j+1}^{(k-1)}
\end{array}
\right..
$$
For the neighborhood of singular point $\mu=0$ we have
\begin{equation}
\varphi_j=\sum\limits_{k=0}^{\infty}C_j^{(k)}\alpha^{1-j-k}\mu^k
\,\mbox{e}^{\displaystyle\alpha t/\mu},
\label{phi_0}
\end{equation}
\begin{equation}
\chi_j=\sum\limits_{k=0}^{\infty}D_j^{(k)}\alpha^{j-k-1}\mu^k
\,\mbox{e}^{\displaystyle-\alpha t/\mu}
\label{chi_0}
\end{equation}
($j=1,...,n$).
Here $\alpha$ is a root of equation
$$
\alpha^n=1,
$$
$C_j^{(k)}$ and $D_j^{(k)}$ are solutions of the systems
$$
\left\{
\begin{array}{l}
C_{j,x_{\mathstrut}}^{(k+1)}=-TC_{j+1}^{(k)}+\sigma_{j,x}C_j^{(k+1)}\\
C_{j,t}^{(k)}+C_j^{(k+1)}=
\mbox{e}^{\sigma_j-T^{-1}\sigma_{j-1}}T^{-1}C_{j-1}^{(k+1)}
\end{array}
\right.,
$$
$$
\left\{
\begin{array}{l}
D_{j,x_{\mathstrut}}^{(k+1)}=T^{-1}D_{j-1}^{(k)}-\sigma_{j,x}D_j^{(k+1)}\\
D_{j,t}^{(k)}-D_j^{(k+1)}=-\mbox{e}^{T\sigma_{j+1}-\sigma_j}TD_{j+1}^{(k+1)}
\end{array}
\right..
$$

One can check by straightforward calculations that the systems determining
coefficients $A_j^{(k)}$, $B_j^{(k)}$, $C_j^{(k)}$ and $D_j^{(k)}$ are
compatible.
The formulas for the expansions in series at singular points neighborhood of
the Lax pairs solutions in nonlocal case seem to be new.
Note that in the case $\mu=\infty$ we have infinitely many types of asymptotic 
behavior.
It follows after substituting expansions (\ref{phi_inf},\ref{chi_inf}) and
(\ref{phi_0},\ref{chi_0}) in Eqs.(\ref{iDT3},\ref{cd},\ref{cf}) that the
solution of the linearization of nonlocal Toda lattice, conserved densities
and currents admit representation
$$
\delta\sigma_j=\sum_{k=0}^{\infty}\delta\sigma_j^{(k,\infty)}\Omega^{-k},\,\,\,
\delta\sigma_j=\sum_{k=0}^{\infty}\delta\sigma_j^{(k,0)}\mu^k,
$$
$$
T=\sum_{k=0}^{\infty}\sum_{j=1}^nT_j^{(k,\infty)}\Omega^{-k},\,\,\,
T=\sum_{k=0}^{\infty}\sum_{j=1}^nT_j^{(k,0)}\mu^k,
$$
$$
X=\sum_{k=0}^{\infty}\sum_{j=1}^nX_j^{(k,\infty)}\Omega^{-k},\,\,\,
X=\sum_{k=0}^{\infty}\sum_{j=1}^nX_j^{(k,0)}\mu^k,
$$
whose coefficients form the infinite hierarchies of infinitesimal symmetries
and conservation laws.
The first few nontrivial coefficients are
$$
\delta\sigma_j^{(1,0)}=\sigma_{j,t}\,,\,\,\,
\delta\sigma_j^{(1,\infty)}=\sigma_{j,x}\,,\,\,\,
\delta\sigma_j^{(2,\infty)}=\sigma_{j,x}^2+
\int\limits^{\,t}\!\left(\mbox{e}^{T\sigma_{j+1}-\sigma_j}+
\mbox{e}^{\sigma_j-T^{-1}\sigma_{j-1}}\right)_xdt.
$$
$$
T_j^{(2,\infty)}=\int\limits^{\,t}\!
\Bigl(\mbox{e}^{T\sigma_j-\sigma_{j-1}}
\int\limits^{\,t}\!
\left(\mbox{e}^{T^2\sigma_{j+1}-T\sigma_j}+
\mbox{e}^{\sigma_{j-1}-T^{-1}\sigma_{j-1}}\right)dt
\Bigr)dt-
\Bigl(\int\limits^{\,t}\!\mbox{e}^{T\sigma_j-\sigma_{j-1}}dt\Bigr)^2,
$$
$$
X_j^{(2,\infty)}=-\mbox{e}^{T\sigma_j-\sigma_{j-1}},
$$

\section{Conclusion}
In this paper the multi--soliton solutions of nonlocal partner of 
two--dimensional Toda lattice were obtained.
Unlike to local case, these solutions depend on infinitely many free 
parameters.
This feature of multi--soliton solutions follows in the frameworks of the 
Darboux transformation technique from analogous property of the solutions of 
the Lax pairs of nonlocal Toda lattice.
We presented also the formulas of the expansions in series on the spectral 
parameter powers of the solutions of the Lax pairs.
These expansions were used to construct the hierarchies of infinitesimal 
symmetries and conservation laws of nonlocal two--dimensional Toda lattice.

\section{ Acknowledgements }
I thank Dr. Heinz Steudel for useful discussions and hospitality.
I thank also Gottlieb Daimler-- und Karl Benz--Stiftung for financial support.

\vfill
\eject
\end{document}